\documentclass[twocolumn,prb,prl,nobibnotes,altaffilletter,amsmath,amssymb,amsfonts]{revtex4}
\usepackage[latin1]{inputenc}
\usepackage{graphicx}
\usepackage{amsmath}
\usepackage{latexsym}
\usepackage{epsfig}

\begin{document}
\title{Generalized Fluctuation-Dissipation Relation and Effective\\
 Temperature upon Heating a Deeply Supercooled Liquid}

\author{Nicoletta Gnan$^1$}
\email{nicoletta.gnan@roma1.infn.it}

\author{Claudio Maggi$^1$}

\author{Giorgio Parisi$^1$}

\author{Francesco Sciortino$^1$}

\affiliation{ $^1$Dipartimento di Fisica, Universit\`a di Roma ``Sapienza'', I-00185, Roma, Italy }

\date{\today}
\begin{abstract}
We show that a generalized fluctuation-dissipation relation applies upon instantaneously increasing the temperature of a deeply supercooled liquid. This has the same two-step shape of the relation found upon cooling the liquid, but with opposite violation, indicating an effective temperature that is lower than bath temperature. We show that the effective temperature exhibits some sensible time-dependence and that it retains its connection with the partitioned phase space visited in aging. We underline the potential relevance of our numerical results for experimental studies of the fluctuation-dissipation relation in glassy systems. 
\end{abstract}

\maketitle

\textit{Introduction} - %%%%%%%%%%%%%%%%%%%%%%%%%%%%%%%%%%%%%%%%%%%%%%%%%%%%%%%%%%%%%%%%%%%%%%%%%%%%%%%%%%%
Supercooled liquids fall out of equilibrium almost inevitably. 
Indeed almost any liquid can be taken to some low-enough temperature (avoiding crystallization) where the structural relaxation processes become extremely slow and the cooling happens to be too quick for the system to equilibrate~\cite{JeppeRev,CavagnaRev}. In this case we observe that the system slowly tries to adapt to the new temperature and exhibits time-dependent thermodynamic properties such as energy or pressure, entering a regime called \emph{physical aging}. In such an off-equilibrium situation the correlation and the response functions characterizing the dynamics of the liquid loose time-translational symmetry and the \textit{fluctuation-dissipation theorem} (FDT) relating the two functions is not supposed to hold anymore~\cite{FDTbooks,VulpianiRev}.

However the interesting fact about glassy systems is that a general fluctuation-dissipation relation (FDR), seems to apply when they fall out-of-equilibrium~\cite{CrisantiRev,CugliandoloRev,KurchanNature,LeuzziBook,Sarracino}. This FDR allows to rigorously define a time-dependent effective temperature 
that is found to reflect the separation of time scales occurring in the relaxation. 
In fact the analytic solution of schematic models~\cite{CugliadoloPSpin,GregorFDT,CrisantiRev, LeuzziBook}, numerical simulations of realistic models of glassy systems~\cite{RdlFDT,KobQCheck,SciortinoTint,BerthierMethod,FDTiso} and few experimental evidences~\cite{GrigeraExp,WangExp,MaggiExp,SchindeleExp}, suggest that, when the system is 
rapidly cooled (quenched) to a low-enough bath temperature $T_{bath}$, the FDR may be written as follows:

\begin{equation} \label{eq:fdr}
T_{bath} \, \partial_{t'} \chi(t,t') = - X(t,t') \, \partial_{t'} \, C(t,t').
\end{equation}

\noindent In Eq.(\ref{eq:fdr}) $C(t,t')=\langle A(t) B(t') \rangle$ is the correlation function of the observables $A$ and $B$ and $\chi(t,t') = \langle A(t) \rangle/\epsilon_{\epsilon \rightarrow 0}$ is the response function to an external perturbation $\epsilon$ applied at the time $t'$ and coupled to the system with an energy contribution $-\epsilon \, B$. $X(t,t')$ on the right-hand side of Eq.(\ref{eq:fdr}) is called \emph{violation factor} and it allows to introduce a time-dependent effective temperature $T_\mathrm{eff} (t,t')= T_{bath}/X(t,t')$. Note that $X=1$ at short time-scales 
(i.e. $(t-t')/t' \ll 1$) where the FDR of Eq.(\ref{eq:fdr}) reduces to the FDT indicating that the fast ``vibrational" dynamics of the molecules is immediately equilibrated at the bath temperature ($T_\mathrm{eff}=T_{bath}$). Differently at long time-scales (i.e. $(t-t')/t' \gg 1$) it is found that $X<1$ suggesting that the slow structural dynamical rearrangements behave as if they were in equilibrium at a temperature $T_\mathrm{eff}=T_{bath}/X$ higher than $T_{bath}$. This is a well-established scenario confirmed theoretically and numerically in the case where the system is taken out of equilibrium by rapidly cooling it down to the glassy phase. Nevertheless slow dynamics and aging effects can be observed in a glassy system without necessarily cooling it. It is possible, for example, increase rapidly the temperature of an equilibrated supercooled liquid or a glass and obtain typical aging effects~\cite{Kovacs,GnanAgingPEL}. 
In this case it is interesting to ask what happens to the FDR and the associated $T_\mathrm{eff}$ since so far no study of this kind has been reported in the literature~\footnote{A theoretical study of the XY model evolving from its ground-state at finite $T$ was performed in Ref.~\cite{BerthierXY}. This displays   $X>1$, hover this violation can be associated to $T_\mathrm{eff}$ only to a qualitative level being absent any clear separation of time-scales.}.

In this work we show, for the first time, that the FDR and the effective temperature concept can be extended to the case where an initially equilibrated deeply supercooled liquid is taken off-equilibrium by suddenly \emph{heating} it and letting it age towards a new equilibrium state. We find that the FDR displays the two slopes observed in quenches but with $X>1$ at large time-scales. This suggests that the slow dynamics behaves as if it were equilibrated at some $T_\mathrm{eff}$ that is \emph{lower} than $T_{bath}$. Moreover we find that $T_\mathrm{eff}$ exhibits a strong dependence on the time elapsed from the instantaneous heating. Finally we show that the connection between $T_\mathrm{eff}$ and the potential energy landscape found in some glassy systems is conserved when the same systems are quickly heated. As argued in the following our results suggest a new route to test experimentally the breakdown of the FDT and measure $T_\mathrm{eff}$.

\textit{Simulation} - %%%%%%%%%%%%%%%%%%%%%%%%%%%%%%%%%%%%%%%%%%%%%%%%%%%%%%%%%%%%%%%%%%%%%%%%%%%%%%%%%%%%% 
We simulate a binary mixture (80:20) of 1000 particles interacting via the Kob-Andersen Lennard-Jones 
(KALJ) potential~\cite{KABLJ1,KABLJ2}. Equilibrium low-temperature configurations are produced by numerically integrating thermostatted (NVT) Newtonian dynamics for times of the order of $10^8$ molecular dynamics (MD) steps for the lowest $T$ investigated\footnote{Unless differently specified standard Lennard-Jones units are used,
the time units in the MD simulations are given by
$1 \, \mathrm{ MD \, step} = 0.005 \, \mathrm{ time \, units}=10^{-14} \, \mathrm{s}$ (in Argon units)}.  
MD simulations are run on high-end graphics processing units (GPUs)~\cite{CUDAmd} that allow for a considerable speed-up of the simulation that typically runs $\sim 20$ times faster on GPUs than on standard CPUs.  Using such a massively-parallel computing approach we produce from $ 5 \times 10^{3}$ to more than $10^4$ independent initial configurations for each low-temperature state from which we perform our numerical heating experiments. The lowest temperature equilibrated is $T=0.41$ at density $\rho = 1.2$, 
where the mode-coupling temperature is $T_\mathrm{MCT}=0.435$~\cite{MCT_KA}. 
For each low-$T$ configuration we follow the off-equilibrium dynamics after an instantaneous temperature increase ($T$ up-jump) using standard Monte Carlo (MC) simulations in the canonical ensemble. 
The choice of MC dynamics allows to employ a zero-field algorithm that can be used for obtaining the integrated response function without actually applying any external perturbation. With this technique, introduced in Ref.~\cite{BerthierMethod}, the response is guaranteed to be free of any non-linear contribution. Moreover this method provides the response as a function of $t'$ (i.e. the instant at which the perturbation is applied) avoiding to run one simulation for each $t'$.

\begin{figure}
\begin{center}
\includegraphics[width=8.6cm]{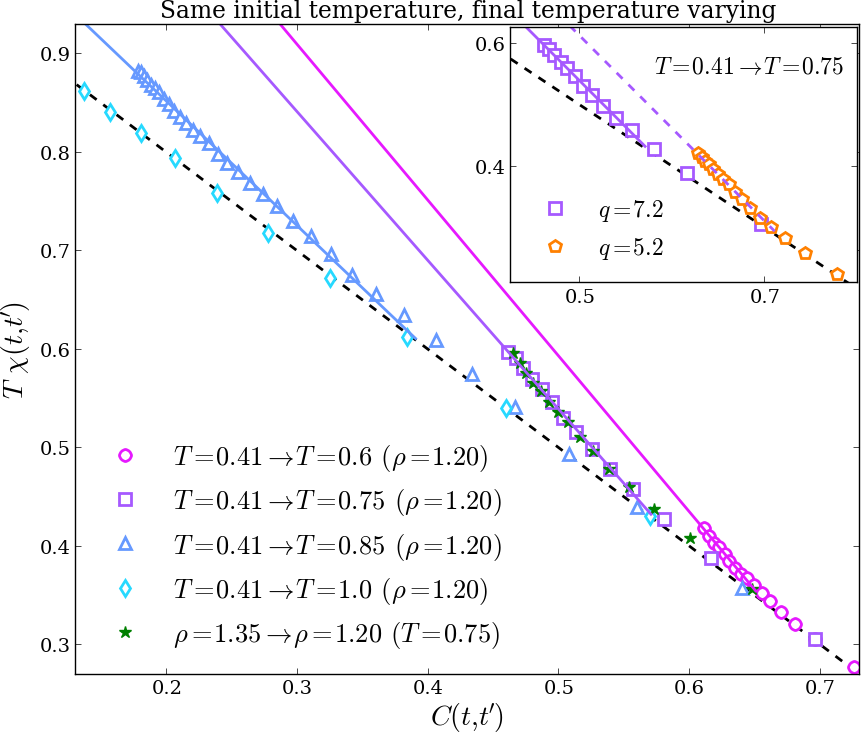}
\caption{
Response versus correlation parametric plots (FD-plots) for several temperature up-jumps as the final (bath) temperature $T_{bath}$
is varied. The time $t'$ varies between $ 6 \times 10^2$ and $t=8 \times 10^3$ MC steps.
A violation of the FDT (dashed line)
characterized by $X>1$ (i.e. $T_\mathrm{eff}<T$) is clearly observed at large time-lags $t-t'$
as evidence by the straight-line fits of the points departing from the FDT. It can be seen how, increasing the final temperature of the jump, the FDT-violation is gradually lost. We show also how two identical FD-plots can be generated by instantaneously increasing the volume (at fied $T$) 
or the temperature (at fixed $\rho$) in analogy with the findings of Ref.~\cite{FDTiso}. 
(Inset) FD-plot in $T$ up-jump for two different wave-vectors. The fitting line of the $k=5.2$ case fits very well also the $k=7.2$ case indicating that $T_\mathrm{eff}$ does not depend on the chosen wave-vector
as found also in quenches~\cite{KobQCheck}. 
}
\label{fig:f1a}
\end{center}
\end{figure}

\begin{figure}
\begin{center}
\includegraphics[width=8.5cm]{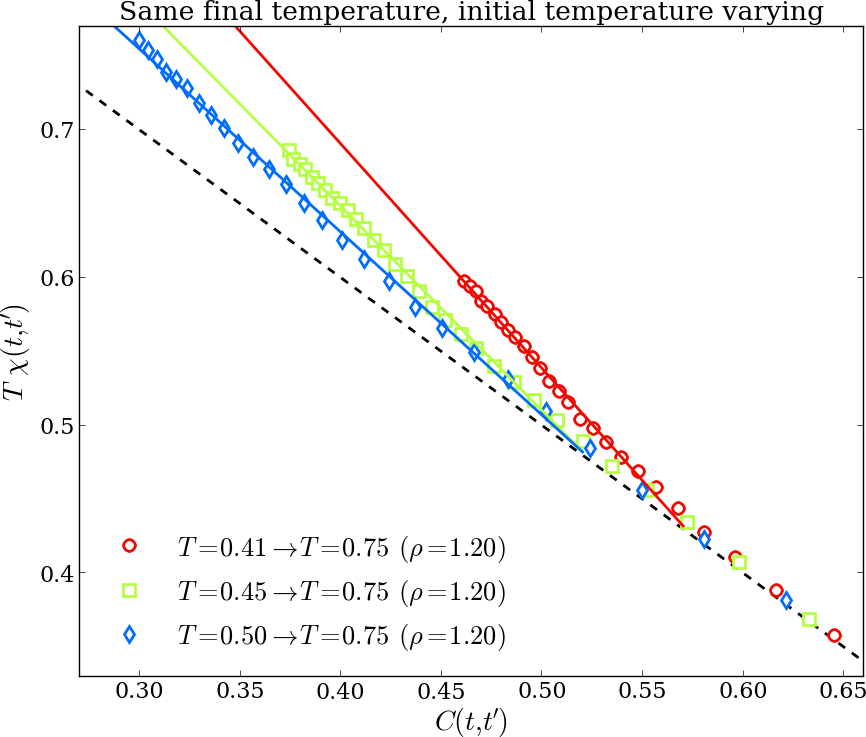}
\caption{
FD-plots for $T$ up-jumps starting with different initial temperatures 
($6 \times 10^2 < t'< 8 \times 10^3 $ MC steps). 
Notice that, as the initial temperature is increased the deviation from the FDT reduces.
}
\label{fig:f1b}
\end{center}
\end{figure}

\textit{Results} - %%%%%%%%%%%%%%%%%%%%%%%%%%%%%%%%%%%%%%%%%%%%%%%%%%%%%%%%%%%%%%%%%%%%%%%%%%%%%%%%%%%%%%%% 
We study the FDR and $T_\mathrm{eff}$ by building the parametric FD-plot reporting $T_{bath} \, \chi(t,t')$ versus $C(t,t')$. In this way the FDT-violation can be immediately visualized and estimated from the parametric curve, being $X = T_{bath}/T_\mathrm{eff}= -T_{bath} \partial \chi(t,t') / \partial C(t,t')$ at large time-scales.  
The variables $A$ and $B$, defining $\chi$ and $C$, are chosen to be  $A_\mathbf{k}(t)=N^{-1} \sum_j \eta_j \exp ( i \mathbf{k} \cdot \mathbf{r}_j(t))$ and $B_\mathbf{k}(t)=N [A_\mathbf{k}(t)+A_{-\mathbf{k}}(t)]$, where the sum is extended to all $N$ particles of the system with instantaneous positions $(\mathbf{r}_1(t),...,\mathbf{r}_N(t))$, and $\eta_j$ is a bimodal random variable with zero mean. 
With this choice the correlation function coincides with the self-intermediate scattering function 
at the wave-vector $\mathbf{k}$.

Figure~\ref{fig:f1a} shows that an evident FDT-violation occurs when $T_{bath}$ is 
instantaneously increased from a very low $T$. 
Note that the FD-plot displays the two slopes already 
found in quenches, but with opposite violation. 
By looking at Fig.~\ref{fig:f1a} it is also clear that 
the violation gradually disappears on increasing the final $T$. 
This is because the relaxation becomes so fast that it is not possible to see clearly, 
neither characterize, a regime where $X \neq 1$ in the FD-plot. 
Differently, if the final $T$ in the up-jump is too close to the initial $T$, 
a very slow relaxation is observed which is not suitable for our study.
Inset of Fig.~\ref{fig:f1a} shows that $X$ does not depend on the wave-vector $\mathbf{k}$ entering in the definition of the correlation and the response, as observed in the case of quenches~\cite{KobQCheck}. 
Moreover we verify that identical FD-plots are obtained if we increase $T$ or if we decrease $\rho$
instantaneously starting from initial states that are connected by the density-scaling equation~\cite{ISOs}(Fig. \ref{fig:f1a}). This confirms that the KALJ system has well-satisfied off-equilibrium scaling properties in heating/expansion protocols as found for cooling/compression procedures~\cite{RdlFDT,FDTiso}.

In Figure~\ref{fig:f1b} we show how the FDR changes as the initial $T$ in the up-jump is changed.
Upon increasing the initial $T$ the deviation from the FDT is weaker making it clear why low-$T$ initial states are necessary for our study. However it is interesting to note that, 
even for an initial $T > T_\mathrm{MCT}$, the deviation from the FDT is still quite evident.
This suggests that the FDT-violation can be observed in wide range of temperatures in the 
supercooled regime above and below $T_\mathrm{MCT}$. 

\begin{figure}
\begin{center}
\includegraphics[width=8.5cm]{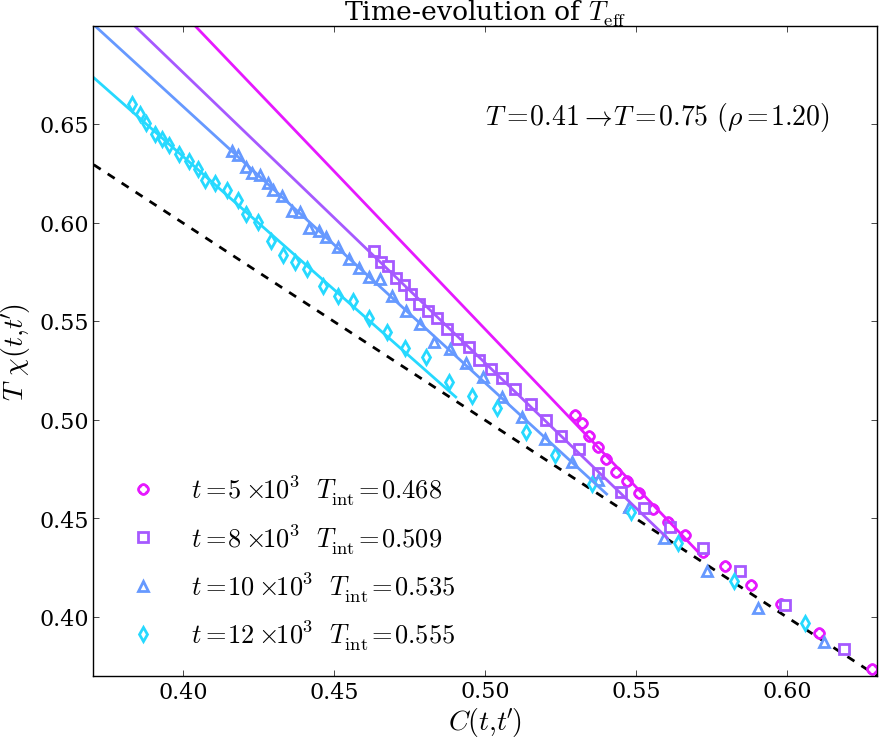}
\caption{
FD-plots for several $T$ up-jumps as the time $t$ increases.
Here $t'$ varies always from $\sim t/10$ to $t$.
The deviation from the FDT (dashed line) 
gradually decreases as $t$ increases suggesting that the slow degrees of 
freedom of system slowly thermalize to the bath temperature.
The straight lines have slope $-T/T_\mathrm{int}(t)$ 
(only the intercept has been adjusted for fitting), 
where $T_\mathrm{int}$ is the internal temperature obtained by using Eq.~(\ref{eq:Tint})~\cite{SciortinoTint} (see text).   
}
\label{fig:f2}
\end{center}
\end{figure}

To further characterize the time-dependence of the FDT-violation we choose as initial equilibrium state $T=0.41$ and as a final state $T=0.75$ ($\rho=1.2$). 
In this jump  the time evolution of the system is slow enough to allow for a characterization 
of the FDR at well-separated aging times.  
Note that the time-scales and the aging times studied here are of the same order of those studied in the quenches~\cite{FDTiso}.
As shown in Fig. \ref{fig:f2} we find that $X$ strongly depends on time 
suggesting that, as the system equilibrates, 
the deviation from the FDT decreases and the $T_\mathrm{eff}$ smoothly approaches the bath temperature $T_\mathrm{bath}$.
This dynamic behavior of the FDR gives the possibility of studying in detail the connection between $T_\mathrm{eff}$ and the configuration space visited out of equilibrium. In the inherent state (IS) formalism\cite{stillinger,jstat} each configuration of the system is associated to the configuration reached by a steepest descent  minimization of  the potential energy. The phase space is partitioned in basins of attraction of different potential energy minima.  In the IS formalism, the free-energy is written as
$F=e_\mathrm{IS}-T_\mathrm{bath} \, s_\mathrm{conf}(e_\mathrm{IS})+f_\mathrm{vib}(e_\mathrm{IS},T_\mathrm{bath})$, where the configurational entropy $s_\mathrm{conf}$  measures the number of distinct minima with energy $e_{IS}$ while the vibrational free energy $f_\mathrm{vib}$   accounts for the contribution to the total
free energy arising from the exploration of the sampled basin volume in configuration space.
Such free energy expression is the IS analog of the TAP free energy\cite{TAP}, where $e_{IS}$  labels the possible TAP states.   In equilibrium,  $\partial{F}/\partial{e_{IS}}=0$, providing a convenient way to calculate
 $\partial s_\mathrm{conf}/\partial e_{IS}$ as 
\begin{equation} \label{eq:sconf}
\frac{\partial s_\mathrm{conf}}{\partial e_\mathrm{IS}} =  T_{bath}^{-1} \left[ 1+ 
\frac{\partial f_\mathrm{vib}(e_\mathrm{IS},T_{bath})}{\partial e_\mathrm{IS}} \right].
\end{equation}

In the IS framework the aging dynamics  is modeled as a walk in configuration space, composed by a fast vibration 
in one minimum and rare jumps among different minima. This approach leads to define an
off-equilibrium extension of the
 free-energy $F$ in which the configurational part is weighted by the internal $T$ of the system $T_\mathrm{int}$ (the analog of the effective temperature)  while the vibrational term is assumed to be in equilibrium with the bath temperature~\cite{SciortinoTint}. This results in   $F=e_\mathrm{IS}-T_\mathrm{int} \, s_\mathrm{conf}(e_\mathrm{IS})+f_\mathrm{vib}(e_\mathrm{IS},T_\mathrm{bath})$.  By analogy with standard thermodynamics $T_\mathrm{int}$ can be obtained by minimizing the free-energy with respect to $e_\mathrm{IS}$:

\begin{equation}\label{eq:eq2}
\frac{\partial F}{\partial e_\mathrm{IS}} = 1-T_\mathrm{int} \frac{\partial s_\mathrm{conf}(e_\mathrm{IS})}{\partial e_\mathrm{IS}} + \frac{\partial f_\mathrm{vib}(e_\mathrm{IS},T_\mathrm{bath})}{\partial e_\mathrm{IS}} = 0
\end{equation}

\noindent  Substituting Eq.~\ref{eq:sconf} in  Eq.~\ref{eq:eq2}
results in a relation among several quantities:
the  $e_{IS}$ value explored during aging,  $T_{bath}$, the equilibrium $T$ at which the 
basin of energy $e_{IS}$ is typically populated ($T_{eq}$) and  $T_{int}$

\begin{equation} \label{eq:Tint}
T_\mathrm{int}(e_\mathrm{IS}) = T_\mathrm{eq}(e_\mathrm{IS})\frac{1+f_\mathrm{vib}(e_\mathrm{IS},T_\mathrm{bath})}
{1+f_\mathrm{vib}(e_\mathrm{IS},T_\mathrm{eq}(e_\mathrm{IS}))}.
\end{equation}

\begin{figure}
\begin{center}
\includegraphics[width=8.5cm]{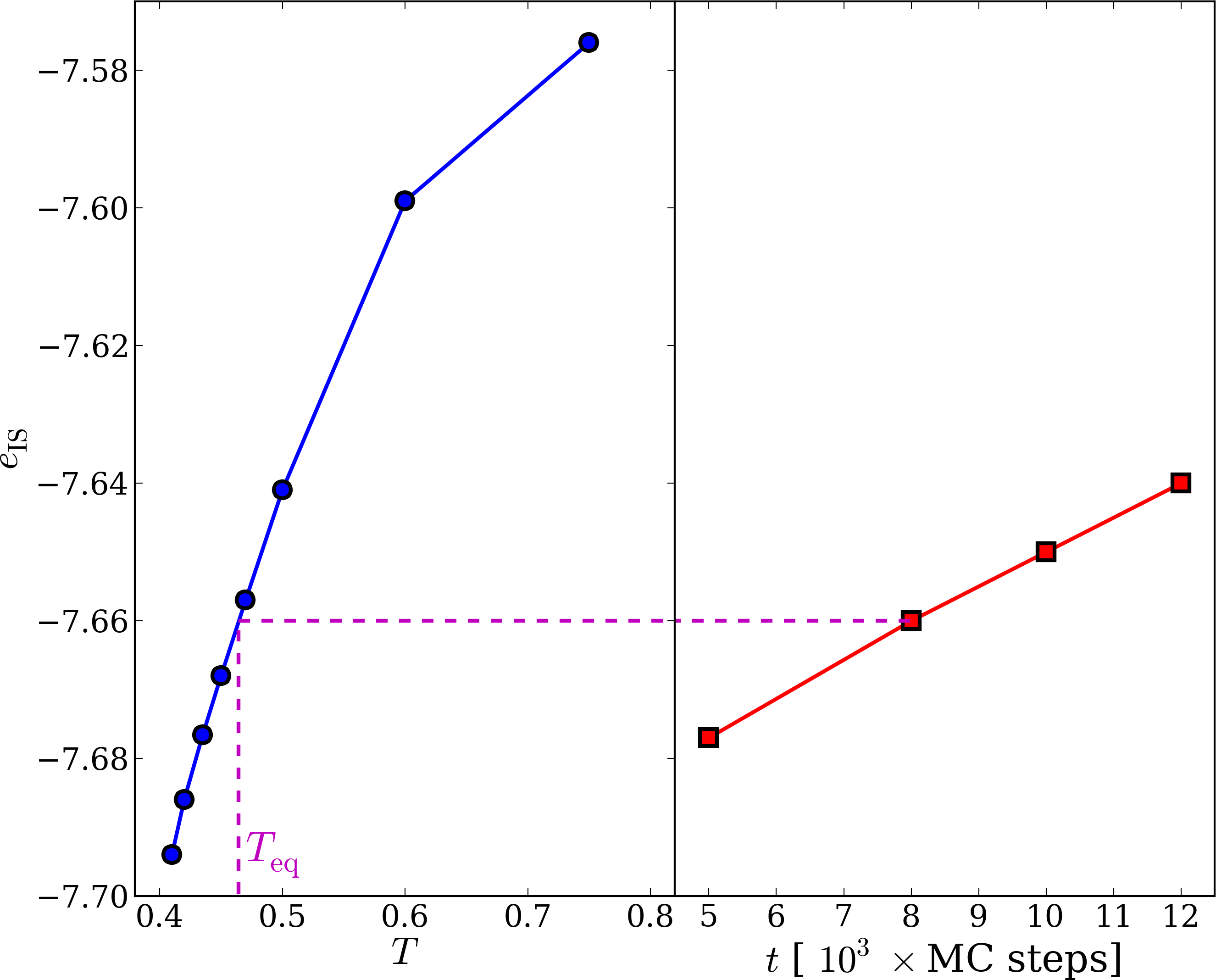}
\caption{
(\textbf{Left panel}): Average inherent structure energy in equilibrium as a function of $T$.
(\textbf{Right panel}): Average inherent structure energy after a $T$ up-jump as a function of time.
The mapping between the equilibrium and the non-equilibrium IS energies allows to estimate the internal temperature that fits very well the FDT-violation (see Fig.~\ref{fig:f2}).
}
\label{fig:f3}
\end{center}
\end{figure}

\noindent  The  graphic method to evaluate   $T_{eq}$ from the knowledge of the $e_{IS}$ explored during the
aging dynamics is reported in Fig.~\ref{fig:f3}. Following Ref.~\cite{SciortinoTint} we determine $f_\mathrm{vib}$ in equilibrium simulations by approximating the IS minima as harmonic 3N-dimensional wells.
 The $T_\mathrm{int}$ value computed via Eq.~(\ref{eq:Tint})  for different aging times can be compared with the slope of the FDR. 
Results displayed in Fig.~\ref{fig:f2} show that, to a very good approximation, $T_\mathrm{eff}(t)=T_\mathrm{int}(t)$, supporting the thermodynamic link implied in the FDR approach. These results suggest that, even in a $T$ up-jump, $T_\mathrm{eff}$ is connected to the dynamics of the underlying IS~\cite{SciortinoTint,FDTiso}. We also note  that the harmonic approximation for $f_\mathrm{vib}$, that is expected to be less accurate as $T$  is increased, still leads to very reasonable results.

\textit{Conclusions} -
We have investigated the off-equilibrium dynamics of a deeply supercooled atomic liquid following an instantaneous increase of the bath temperature. Relying on GPU-implemented numerical simulations we have focused on the FDR and the effective temperature that can be measured out of equilibrium. We have shown that FDR is characterized by the typical two-step shape associated with the time-scale separation of glassy dynamics. The violation factor is found to be greater than one giving a $T_\mathrm{eff}$ lower than the $T_\mathrm{bath}$ temperature  that is the opposite of what is found upon cooling a liquid in the glassy phase. We have identified the most suitable equilibrium states among which the temperature up-jump has to be performed to observe and characterize in detail the FDR. The effective temperature measured displays a clear time-dependence. Further analysis indicates that $T_\mathrm{eff}$ coincides to a good approximation with the internal temperature obtained by a extended (off-equilibrium) thermodynamic framework based on the potential energy phase-space decomposition.

Our results suggest a new route to follow in the experimental study of the FDR. In these challenging experiments one has indeed to measure independently response and correlation as the temperature is quickly changed. The temperature change has to be very fast with respect to the relaxation time of the system. One possibility suggested by our results is to try to perform an heating experiment among equilibrium liquid states that can be quite easily prepared experimentally. These experiments will possibly take advantage from the fact that rapidly heating the liquid might be much easier than cooling if the heating is performed by using, for example, some laser-based techniques~\cite{LaserHeating1,LaserHeating2}. It could be,  quite unexpectedly, that the robust concepts of fluctuation dissipation relation and effective temperature, that were conceived for characterizing the fast cooling of a liquid, are instead more easily accessible in a heating experiment.

%\clearpage
NG and FS acknowledge support from ERC-$226207$-PATCHYCOLLOIDS.


\begin{thebibliography}{99}

\bibitem{JeppeRev} J. C. Dyre, Rev. Mod. Phys. \textbf{78}, 953 (2006)

\bibitem{CavagnaRev} A. Cavagna, Phys. Rep. \textbf{476}, 51 (2009)

\bibitem{FDTbooks} J.-P. Hansen and I.R. Mc-
Donald. Theory of Simple Liquids, 3rd edition Chap. 7.
(Academic Press 2006), 
D. Chandler. Introduction to Modern Statistical
Mechanics (Oxford University Press 1987).

\bibitem{VulpianiRev} U. M. B. Marconi, A. Puglisi, L. Rondoni, et al., Phys. Rep. \textbf{461}, 111 (2008)

\bibitem{CrisantiRev} A. Crisanti and F. Ritort, J. Phys. A: Math. Gen. \textbf{36} R181, (2003)

\bibitem{CugliandoloRev} L. F. Cugliandolo,  J. Phys. A: Math. Theor. \textbf{44}, 483001 (2011)

\bibitem{KurchanNature} J. Kurchan, Nature \textbf{433}, 222 (2005)

\bibitem{LeuzziBook} L. Leuzzi, T.M. Nieuwenhuizen, \textit{Thermodynamics of the Glassy State}, Taylor \& Francis (2007)

\bibitem{Sarracino} E. Lippiello, F. Corberi, A. Sarracino et al.,  Phys. Rev. E \textbf{78}, 041120 (2008)

\bibitem{CugliadoloPSpin} L. F. Cugliandolo and J. Kurchan, Phys. Rev. Lett. \textbf{71}, 173 (1993)

\bibitem{GregorFDT} G. Diezemann, J. Chem. Phys. \textbf{123}, 204510 (2005)

\bibitem{RdlFDT} R. Di Leonardo, L. Angelani, G. Parisi, et al., Phys. Rev. Lett. \textbf{84}, 6054 (2000).

\bibitem{KobQCheck} W. Kob and J. L. Barrat, Eur. Phys. J. B \textbf{13} 319 (2000)

\bibitem{SciortinoTint} F. Sciortino and P. Tartaglia,  Phys. Rev. Lett. \textbf{86}, 107 (2001)

\bibitem{FDTiso} N. Gnan, C. Maggi, T. B. Schroeder et al., Phys. Rev. Lett. \textbf{104}, 125902 (2010)

\bibitem{BerthierMethod} L. Berthier, Phys. Rev. Lett. \textbf{98}, 220601 (2007).

\bibitem{GrigeraExp} T. Grigera and N. E. Israeloff, Phys. Rev. Lett. \textbf{83}, 5038 (1999)

\bibitem{WangExp} P. Wang, C. Song, and H. A. Makse, Nature Phys. \textbf{2}, 526 (2006)

\bibitem{MaggiExp} C. Maggi, R. Di Leonardo, J. C. Dyre, et al., Phys. Rev. B \textbf{81}, 104201 (2010) 

\bibitem{SchindeleExp} J. Schindele, A. Reiser, and C. Enss, Phys. Rev. Lett. \textbf{107}, 095701 (2011)

\bibitem{Kovacs} A. J. Kovacs, Adv. Polym. Sci. \textbf{3}, 394 (1963)

\bibitem{GnanAgingPEL} C. Rehwald, N. Gnan, A. Heuer, et al., Phys. Rev. E \textbf{82}, 021503 (2010)

\bibitem{KABLJ1} W. Kob and H. C. Andersen., Phys. Rev. Lett. \textbf{73}, 1376 (1994)

\bibitem{KABLJ2} L. Berthier and W. Kob, J. Phys.: Condens. Matter \textbf{19}, 205130 (2007)

\bibitem{CUDAmd} J. A. Anderson, C. D. Lorenz, A. Travesseta, J. Comput. Phys., \textbf{227}, 5342 (2008)

\bibitem{MCT_KA} W. Kob, H. C. Andersen, Phys. Rev. E, \textbf{51}, 4626 (1995),
 W. Kob, H. C. Andersen, Phys. Rev. E, \textbf{52}, 4134 (1995)

\bibitem{ISOs} 
N. Gnan, T. B. Schroeder, U. R. Pedersen, et al. J. Chem. Phys. \textbf{131}, 234504 (2009) 
T. B. Schroeder, N. Gnan, U. R. Pedersen, et al. J. Chem. Phys. \textbf{134} , 164505 (2011)
D. Gundermann, U. R. Pedersen, T. Hecksher, et al., Nat. Phys. \textbf{7}, 816 (2011).  

\bibitem{stillinger} F. H. Stillinger  and T. A. Weber, Phys. Rev. A \textbf{25} 978 (1982). 
	
\bibitem{jstat}	 F. Sciortino,
J. Stat. Mech. \textbf{050515},  (2005).

\bibitem{TAP} D. J. Thouless, P.W. Anderson, and R. G. Palmer, Philos.
Mag. \textbf{35}, 593 (1977); A. Crisanti, H. Horner, and H. J.
Sommers, Z. Phys. B \textbf{92}, 257 (1993).

\bibitem{BerthierXY} L. Berthier, P. C W Holdsworth and M. Sellitto,  J. Phys. A: Math. Gen. \textbf{34}, 1805 (2001)

\bibitem{LaserHeating1} S. Ansell, S. Krishnan, J. K. R. Weber, Phys. Rev. Lett. \textbf{78}, 464 (1997)

\bibitem{LaserHeating2} S. Ansell, S. Krishnan, J. J. Felten, J. Phys.: Condens. Matter \textbf{10}, L73 (1998)
 	



\end{thebibliography}
\end{document}